\documentclass[conference]{IEEEtran}
\IEEEoverridecommandlockouts
\usepackage{amsmath,amssymb,amsfonts}
\usepackage{graphicx}
\usepackage{amsfonts}
\usepackage{multirow}
\usepackage{caption}

\def\BibTeX{{\rm B\kern-.05em{\sc i\kern-.025em b}\kern-.08em
    T\kern-.1667em\lower.7ex\hbox{E}\kern-.125emX}}
    
    \usepackage[pscoord]{eso-pic}

\begin{document}

\title{ Dealing with complex routing requirements using an MCDM based approach}

\author{\IEEEauthorblockN{ Mohamed Redha BOUAKOUK}
\IEEEauthorblockA{\textit{LSI Laboratory  } 
\textit{USTHB University} \\
Algiers, Algeria \\
 { \scriptsize mohamedredha.bouakouk@gmail.com }
}
\and
\IEEEauthorblockN{ Abdelkrim ABDELLI}
\IEEEauthorblockA{\textit{LSI Laboratory} 
\textit{USTHB University}\\
Algiers, Algeria \\
 { \scriptsize aabdelli@usthb.dz }
}
\and
\IEEEauthorblockN{Lynda MOKDAD}
\IEEEauthorblockA{\textit{Univ Paris Est Creteil,} \\
\textit{LACL,}
  F-94010 Creteil \\
{ \scriptsize lynda.mokdad@u-pec.fr }
}

\and
\IEEEauthorblockN{Jalel Ben Othman}
\IEEEauthorblockA{\textit{Univ Paris-Saclay,} \\
\textit{ L2S,}
  F-94010 Creteil \\
 { \scriptsize jalel.benothman@l2s.centralesupelec.fr }
}

}
\maketitle

\begin{abstract} The last decade has witnessed an ever-growing user demand for a better QoS (Quality Of Service) and the fast growth of connected devices still put high pressure on the legacy network infrastructures. To improve network performances, better manage the resources and have a greater control over traffic transmission, intelligent routing procedures are increasingly demanded. 
Modern applications in the dynamic context of new emerging networks have their own routing requirements, in terms of set of metrics to consider, their importance and thresholds to respect. 
The objective of this work is to design an approach based on MCDM  (Multi-Criteria Decision Making) to decide complex routing problems when assuming threshold constraints on metrics. We give the mathematical framework to capture such requirements and to decide the routing. A case study is presented to advocate the benefit of using our approach.

\end{abstract}

\begin{IEEEkeywords}
MCDM, Routing, QoS, Requirements, Metrics.
\end{IEEEkeywords}

\section{Introduction}

From a luxury to a necessity, the Internet is moving to a new era.
Indeed,  the number of connected devices  is continuously escalating due to the advent of new emerging technologies such as Internet of things (IoT), MANETS, Device-to-Device Communication (D2D), Fog Computing, Vehicular Ad hoc Networks (VANET), 5G networks, etc \cite{Bou2020}. 
By 2023, it is expected to reach more than 29.3 billion networked devices, thus demanding the increase of the fixed and mobile connection speeds to reach respectively 110.4 Mbps and 43.9 Mbps in 2023. Therefore, the legacy infrastructure can hardly deal with the exponential insistence of the emerging digital era, nor  guarantee the QoS  and the specific requirements of new trendy applications while using the resources effectively and efficiently. 
Beside the physical infrastructure limitations, the routing procedures appear to be no longer able to respond to the increasing demand in terms of requirements (Higher data throughput, lower latency, and less energy consumption).   

For example, in D2D communications which is one of the key components of 5G network infrastructure, we need to monitor the E2ED (End to End Delay) during routing to avoid data congestion at device level. Besides, routing in such networks should consider the mobility of devices unlike in orthodox cellular networking to improve the QoS \cite{p18}. 
In other respects, the RPL protocol has been standardized recently to work in low-power and lossy networks (LLN) using IPV6 as a main part of the IoT infrastructure \cite{Bou2021}.  In RPL, a node selects the routing parent according to its objective function (OF) and accordingly constructs  a destination oriented directed acyclic graph.  Although RPL introduces a flexible routing procedure,  there exists no explicit method to combine different metrics in the OF for selecting the optimum route. Therefore, depending on the specific application requirements, new OFs have to be designed to capture different metrics such as the expected transmission count, hop count, the delay, link failure,  and the energy consumption. Moreover, RPL needs to be improved to cope with congestion and heavy traffic load to reduce the impact on QoS in IoT applications \cite{p14}.  


In the context of MANETs, optimal route selection plays a vital role in improving the overall QoS and the network lifetime, as nodes are energy constrained and the topology is dynamic due to the high mobility of nodes. The complexity of the routing escalates with the number of nodes as the number of possible paths to reach a destination is expanded. Traditional routing protocols in MANEts are usually classified into reactive (like in AODV and DSR) and proactive (like in OLSR). Proactive ones are specifically designed to reduce the E2ED while the reactive ones are considered to reduce the control message overhead by employing a single routing metric. 
To be able to cope with specif application requirements, the research is exploring the use of  multiple metrics by differentiating their importance, such as link quality, bandwidth and available energy during the routing decision. More specifically, adjusting the routing decision according to desired E2ED, reliability, jitter and energy thresholds is becoming paramount.


To achieve the full potential of WSN (Wireless sensor Networks) through the Internet, it is recommended to deploy IP based network complying with the current and existing standards. This calls to redesign  conventional routing algorithms to cope with  some conditions or threshold values.  Moreover, modern multimedia streaming applications  have stringent QoS requirements of throughput, delay and packet loss that should be expressed as hard or soft constraints \cite{Abd2017}. For example, the data rate of H.264 should varies between 64 kbps and 240 Mbps depending on the QoS level.  
Unlike in WSN, in  Wireless Body Area Network (WBAN), there are additional routing issues to deal with. For example, it is more important to consider routes that reduce the energy consumption, the packet loss and the transmission delay while taking into account  the  devices temperature and the human posture \cite{p10}.

Consequently, to be capable to capture complex user application requirements in the routing problem decision, we explore in this paper an approach based on using  MCDM  methods. Indeed, applying MCDM allows to find optimal results in complex scenarios including various indicators, conflicting objectives and criteria.  When dealing with routing decisions, MCDM can flexibly cope with various network scenarios and application requirements.  However, traditional MCDM approaches are unable to capture threshold values when expressed on the metrics. To the best of our knowledge, differentiating between the importance of the considered criteria. is the only aspect that has been addressed in the existing  MCDM based routing solutions. To deal with complex requirements like specifying values constraints to meet during the decision making process, a new MCDM method called ISOCOV (Ideal SOlution for COnstraints on Values) has been put forward recently \cite{Isocov1, Isocov2}. We explicit the different procedural steps on applying ISOCOV on a routing problem with threshold requirements when assuming the latter either as mandatory or as soft. A case study is presented to advocate the capabilities of our approach comparing to classical ones.   

The remainder of this paper is organized as follows. Section II discusses the related works. In section III, we show how to model user application requirements in routing problems using ISOCOV.  
Section IV shows how the routing problem is decided mathematically.
In Section V, we illustrate the approach through a case study.
Section VI concludes the paper.

\section{Related works}

In the literature, several MCDM methods have been developed to solve different multi-criteria problems, as for instance: AHP \cite{AHP},  TOPSIS \cite{TOPSIS}, VIKOR \cite{VIKOR}, SAW \cite{SAW}  etc.
Almost all the MCDM methods are assuming that the optimal values of the performance ratings are obtained by considering the minimum or the maximum of the range (ideal values). However, in practice, we may be interested to specify these ideal values as an interval somewhere in between the range of the performance ratings, or simply to express a preference for a specific interval of values, which are not necessarily the optimum.
So far, only very few recent MCDM methods have proposed to extend the decision making framework to cope with such requirements, we can quote:  RIM (\textit{Reference Ideal Method}) \cite{RIM}, TOPSIS-Star \cite{Serai2018}, and ISOCOV   \cite{Isocov1, Isocov2}. 
RIM and TOPSIS-Star assume that the optimal values must be within the specified intervals thus providing a partial ranking only for alternatives whose ratings are outside the ideal intervals.  To offer more possibilities, ISOCOV has been introduced to integrate the preferences of the Decision Maker (DM), as value constraints in form of intervals with the possibility  to specify the constraints as mandatory or as soft ones. ISOCOV thus makes it possible to provide a complete ranking of all the alternatives by integrating the performance ratings as well as their degrees of constraint satisfaction in the decision making which is more suitable in the case of  routing problems. To the best of our knowledge, all the existing solutions  that used MCDM methods to improve routing, have only dealt with route selection problem when assuming different metrics with differentiated weights, without considering user application preferences in terms of specific metrics values to achieve. We review in the following the most recent works in this topic. 


In next generation networks, like 5G, and IoT, each connected network is different and requires to be evaluated in terms of specific metrics like coverage area, available bandwidth, cost, etc.
In \cite{p5}, the authors compare the use of several MCDM methods during the handover between different networks. The MCDM is applied to decide which intermediate network should be considered in the routing process. In \cite{p18}, the authors explore an MCDM based routing approach in  5G D2D communications. The MCDM problem consists in finding the best route that optimizes energy, mobility, queue length, and link quality-aware. 
Moreover,  the use of MCDM to improve security, pervasive reliability and QoS in hostile delay tolerant network routing is explored in \cite{p2}. The effectiveness of the solution against different security metrics are assessed in the presence of bad-mouthing, good-mouthing, and selfish attacks.
In \cite{p17}, the authors propose a congestion-aware routing protocol (CoAR) that relies on using MCDM to select the  best alternative parent node within the congestion by combining different routing metrics. 
In the  same regard, RPL is extended in \cite{p14} with an MCDM scheme to improve the QoS of routing. They adopt VIKOR  to select the best parent in the path according to the  average energy consumption, the E2ED, the Packet Delivery Ratio (PDR) and the throughput.

In the context of WSN, the sensed data may reach sink node through multiple hops event driven routes. To select the best route according to different metrics, the authors in \cite{p6}, propose an MCDM based routing scheme. The weight of the metrics are determined using an adaptive learning method.
Moreover, the authors  in \cite{p10} propose an enhanced mobility and temperature-aware routing protocol based on MCDM to route data in  WBAN. They use AHP and SAW  to assign suitable weight factors and to select the next hop in he route according to multiple metrics.  The goal is to determine the optimal route with the lowest devices temperature, transmission delay and data loss caused by human posture. 


The  selection of the optimal route based upon QoS requirements is more than desirable in a completely dynamic environment like MANET. 
To deal with such an uncertain environment, the authors in \cite{p16} extends AOMDV by using a Fuzzy TOPSIS based method to select the optimal path among different available routes according to four QoS parameters which are: reliability, bandwidth jitter and delay. In a similar work \cite{p9},  AODV and OLSR are extended with an MCDM scheme to select a route according to multiple metrics. 
The MCDM method is also used to determine the weights of the latter. The same authors investigate in \cite{p3} the use of MCDM method to improve geographical routing to accommodate multiple metrics in a logical way. They propose a new dynamic priority scheme to select the most suitable next hop by differentiating the priorities according to the network environment.
Moreover, in \cite{p13} the authors have considered  nodes mobility, contention window size, and link quality to select the optimal route using TOPSIS. The latter is also applied to determine the importance of each metric. 
Moreover, in opportunistic networks, the selection of the route depends on different metrics which are highly correlated. Add to this, network dynamicity  affects the importance of each metric as well as the levels of its correlation with other metrics. 
Within this context, the authors in \cite{p1}  propose an MCDM based approach to normalize a set of metrics  to evaluate different routes  according to special requirements that are specific to the opportunistic context.

\section{Modelling an application's constraints routing problem using ISOCOV}
 
In this section, we identify the different user application requirements that can be specified in a routing problem. Then we present our approach based on ISOCOV to capture them  to make the routing decision. 

\subsection{Modelling routes QoS performances}
 
Let us consider a set of $m$ alternative routes $a_{i}$ going from a source node $NS$ to a destination node $ND$. When $NS$ has to send a packet to $ND$, it needs to  select the best route  in terms of QoS performances to route the packet according to the QoS requirements specified in the packet header. These requirements are generally set by the application that has generated the packet.
The $m$ possible routes, are characterized by different QoS metrics, that assess the performances of the given route. These QoS metrics can relate, for instance, to the throughput, the E2ED, the hop count, the PDR, the jitter, etc
We assume the existence of $n$ QoS metrics, noted $Q_j$ to evaluate the $m$ routes.
Moreover, a QoS metric can either denote a \textit{cost} or a \textit{benefit}. A cost metric has to be minimized, whereas a benefit metric is to be maximised. For example, the throughput is a benefit metric, while the E2ED, the hop count and the PDR are cost metrics.
The performances ratings of the $m$ routes relatively to the $n$ QoS metrics can be encoded by a decision matrix, noted $D$, of dimensions $(m,n)$. The value $d_{ij}$ thus denotes the performance of the route  $a_{i}$  relatively to the metric $Q_j$. 

\subsection{ Modelling complex routing requirements} 

We show in the sequel how to model the QoS requirements needed by the packet during its route, and which are fixed beforehand by the user application.
In MCDM methods, the decision making assumes a weight vector denoting the importance of each criterion relatively to the others. In routing problem modelling, the criteria refer to QoS metrics. 
We note $w_{j}$ the weight of the metric $Q_{j}$, that quantifies its importance relatively to other metrics, according to user application requirements. 
If a metric $Q_{j}$ is not required during the routing process, then we have $w_{j}=0$. Moreover, we assume that the $n$ weights are known and normalised, namely:\\
$\sum_{j=1..n}  w_{j}= 1$ such that $w_{j}\in [0,1]$.

By following a calculation process given by the MCDM method, each route will obtain an aggregating score measuring its global performances regarding all the requirements. These scores are then used to rank the $m$ routes and hence to select the one that better matches user application requirements. Changing the weights of the metrics will impact the performances of the routes and therefore the final scores and the rankings.
All the traditional and popular MCDM methods, like TOPSIS and VIKOR
and their different extensions offer the ability to express the importance of the different criteria as normalized values when they are known, or by providing a formal framework to quantify them  when they are expressed as fuzzy numbers.   
Applying such algorithms on routing problems will decide which route is the best according to application preferences in terms of metrics to be promoted.
However, when value constraints are expressed on the performance ratings of the different routes,  as additional requirements, the majority of MCDM methods are enable to cope efficiently with that.
In essence, if a user application requires a route such that the E2ED should be under a given threshold, the only way to handle such a request is to discard all the alternatives that do not fulfil this requirement, and proceed to rank only routes that satisfy all the constraints \cite{Serai2017}.  
Unlike other MCDM methods which are not designed to incorporate value constraints within the decision making process, ISOCOV is a new approach that has been introduced to cope with such requirements in a flexible manner.  ISOCOV is designed to offer a better solution towards this problematic by defining a mathematical framework to specify the value constraints, as additional requirements. 
Moreover, ISOCOV defines flexible formulae to compute the aggregation scores by considering the nature of the constraints, whether they are hard or soft. Concretely, according to our case study: If we deal with \textit{hard constraints} this means that the  value constraints on QoS metrics are mandatory. In this case, alternative routes  are distinguished into two groups. Those that are meeting all the constraints and are ranked on the top relatively to those that are missing the constraints which follow in the ranking. Such a hard requirement can be observed, for example,  in Non Delay Tolerant routing protocols, and multimedia applications which need, for instance, that the E2ED or the jitter must be under a given threshold, or within a given interval. In this case, ISOCOV will promote as best routes all the alternative that are meeting all the constraints and rank them according to their performances. The second set which contains ranked routes that are not meeting all the constraints, is provided as an alternative to be considered if the first set is empty of if its routes are unavailable or off.        
On the other hand, if we deal with  soft constraints, this means that the routing process follows a best effort procedure.
The router strives to promote the best route in terms of performances without any guarantee to meet all the application constraints. However, routes that satisfy all the constraints have more chances to be promoted than the rest of alternatives.
Concretely, ISOCOV considers the following additional parameters to encode user application constraints.

\begin{itemize}
    \item $Hard$: is a boolean variable that takes the value $true$ if the constraints are mandatory, and $false$ if they are soft.
    \item  For each metric $Q_{j}$, its value constraints are expressed as an interval: $CV_{j}=[A_{j},B_{j}]$, such that $[A_{j},B_{j}] \subseteq [d_{j}^{min},  d_{j}^{max}] $, and we have:
$d_{j}^{min}=MIN^{m}_{i=1} \{ d_{ij} \} $ and $d_{j}^{max}=MAX^{m}_{i=1}  \{ d_{ij} \}$  be respectively the minimum and the maximum performance ratings for the metric $Q_{j}$.
\end{itemize}

\textbf{Remark: }In the sequel we assume the following notations:  $CV_{j}=[\circ, \circ ]=  [d_{j}^{min},d_{j}^{max}  ], $\quad
$CV_{j}=[A_{j}, \circ ]=[A_{j},d_{j}^{max}  ] $, \quad
$CV_{j}=[\circ, B_{j}]= [d_{j}^{min},B_{j} ] $.

\section{Ranking process with ISOCOV }
In this section, we give the different steps of the ISOCOV computational framework to compute the score associated with each  route. The $m$ obtained scores are then used to rank the $m$  related alternative routes.

\subsection{Computing the degrees of constraints satisfaction}
The first step of the computation process consists in encoding the
closeness to meet the value constraints as coefficients while varying the routes and the metrics. Therefore, for each route $(a_i)$ and each metric $Q_{j}$, we compute the degree of closeness of the performance of the route $(a_i)$ relatively to the constraint $CV_{j}$ defined on the metric $Q_{j}$. 
 Each \textit{degree of constraint satisfaction}, noted $ f_{ij}$, which takes its values within $(0,1]$, is computed using the formulae (1) and (2) according to the nature of the metric $Q_{j}$. 
If $Q_{j}$ is a benefit

$ f_{ij} =$
 \begin{equation}
\left 
\{ 
\begin{array}{l}
1  \qquad \qquad \qquad \qquad \qquad \qquad \qquad  if d_{ij} \in  [A_{j},B_{j}]  \\
1- \frac{A_{j}-d_{ij} }{ MAX(A_{j}-d_{j}^{min}, \quad d_{j}^{max}-B_{j} )+1} \quad     if d_{ij}  \in  [d_{j}^{min},A_{j}) \\
1- \frac{d_{ij}-B_{j}}{MAX(A_{j}-d_{j}^{min}, \quad  d_{j}^{max}-B_{j} )+1}  \quad  if d_{ij} \in  (B_{j},d_{j}^{max}] 

\end{array}
\right. 
\end{equation}

If $Q_{j}$ is a cost

$ f_{ij} =$
\begin{equation}
\left 
\{ 
\begin{array}{l}
\frac{1 }{ MAX(A_{j}-d_{j}^{min}, \qquad d_{j}^{max}-B_{j} )+1}   \quad  if d_{ij} \in  [A_{j},B_{j}]  \\
\frac{A_{j}-d_{ij} }{ MAX(A_{j}-d_{j}^{min}, \qquad d_{j}^{max}-B_{j} )} \quad    if d_{ij}  \in  [d_{j}^{min},A_{j}) \\
\frac{d_{ij}-B_{j}}{MAX(A_{j}-d_{j}^{min}, \quad  d_{j}^{max}-B_{j} )}  \qquad  if d_{ij} \in  (B_{j},d_{j}^{max}] 

\end{array}
\right. 
\end{equation}

In essence, if $Q_{j}$ is a benefit and the route $(a_i)$ is meeting the related constraints $CV_{j}$, it obtains a maximum value equal to "1". Otherwise, the value of $f_{ij}<1$ will denote the closeness to meet the constraints $CV_{j}$. More the route is close to meet the latter, more $f_{ij}$ will be close to 1. 
On the other hand, in case $Q_{j}$ is a cost, the route $(a_i)$ gets the minimum score if its satisfies the constraints $CV_{j}$.
More $f_{ij}$ is close to 1, less the route is close to meet the constraints. The different obtained scores encoding the degrees of constraints satisfaction are encoded as a matrix $F(m,n)$. This matrix is used in the normalization process of the performance ratings.

\subsection{Normalization of the performances ratings}

The different metrics are introduced to rate the performances of the different routes without assuming necessarily the same data type or the same range.  To be able to compute the final performance scores we need first to normalize  the matrix $D$. This normalisation process integrates the weights of the different metrics, as well as  the degrees of constraints satisfaction to proportionate the importance of each performance accordingly. The normalized and weighted matrix, thus obtained, is noted $P$ and is computed as follows: \medskip 

\textbf{Step 1}: Normalize the matrix $D$ using the vector normalization formula:
\begin{equation} N_{ij}=\frac {d_{ij}} {\sqrt{\sum_{i=1}^{m}(d_{ij})^{2}}}  \quad i=1..m, \quad  j=1..n \end{equation}

\textbf{Step 2:} Compute the product of the normalized data encoded in $N$ and the respective metric weights $w_{j}$ and the degrees of constraints satisfaction $f_{ij}$: 
\begin{equation}
p_{ij}=N_{ij} \times w_{j} \times f_{ij}  \qquad i=1..m, \quad  j=1..n
\end{equation}

\subsection{Computing the closeness coefficients}

After normalizing and weighting the data, the next and last step consists in aggregating the latter to calculate a score associated with each route. To this aim, we need first to compute two vectors, the PIS (Positive Ideal Solution) and the (Negative Ideal Solution) NIS, as follows:\smallskip

\textbf{Step 3:} Compute the PIS  and the  NIS, noted respectively $R^{+}$ and $R^{-}$, according to the nature of each metric:
\newline
If $Q_{j}$ is a benefit, then
\begin{equation}  R^{+}_{j}=MAX^{m}_{i=1} \{ p_{ij} \}. \quad R^{-}_{j}=MIN^{m}_{i=1}\{p_{ij}\} \end{equation}  
If $Q_{j}$ is a cost, then 
\begin{equation}
R^{+}_{j}=MIN^{m}_{i=1} \{p_{ij}\}. \quad R^{-}_{j}=MAX^{m}_{i=1} \{p_{ij} \} \end{equation} 
\textbf{Step 4:} Then, we compute the euclidean distance  between each route \textit{$i = 1,2,..,m$  } and both the PIS and the NIS, as follows:
 
\quad \textbf{Euclidean distance from the PIS}: 
  \begin{equation} S_{i}^{+}=\sqrt{\sum_{j=1}^{n} ( R^{+}_{j}-p_{ij} )^{2} } \end{equation}  
\quad \textbf{Euclidean distance  from the NIS}:
 \begin{equation} S_{i}^{-}=\sqrt{\sum_{j=1}^{n} (p_{ij}- R^{-}_{j} )^{2} }  \end{equation} 
\textbf{Step 5:} According to the nature of the constraints, ISOCOV computes for each route its aggregation score, as follows:

If $Hard=true$ then
\begin{equation} 
 C_{i}=\frac{S_{i}^{-}}{S_{i}^{-}+S_{i}^{+}} \quad if V(i)=1 \\
\end{equation} 
 \begin{equation} 
C_{i}= \frac{S_{i}^{-}}{S_{i}^{-}+S_{i}^{+}}-1  \quad  if V(i)=0 \\
 \end{equation}
 
If $Hard=false$ 
\begin{equation} C_{i}=\frac{S_{i}^{-}}{S_{i}^{-}+S_{i}^{+}}   \end{equation} 
 Such that, $V(i)$ is a binary variable, associated with the route $a_i$.
\begin{tabular}{cc}
 $V(i)=1$  &  if $\forall j=1..n \quad d_{ij}\in   [A_{j},B_{j}] $ \\
 $V(i)=0$  & otherwise \\
\end{tabular}

In other terms, $V(i)=1$ if the route $(i)$ meets all the  value constraints defined on the $n$ metrics.

In case the constraints are hard, routes that are meeting all the values constraints ($V(i)=1$) obtain an aggregation score within the interval $[0,1]$. The other routes ($V(i)=0$) are granted  scores within $[-1,0]$. However, if we are dealing with soft constraints all the routes get their scores within the interval $[0,1]$ whatever they meet all the constraints or not. The final performance score of each route $C_i$ integrates its global normalized performance ratings weighted by the importance of the $n$ metrics, and the constraints satisfaction degrees. \smallskip

\textbf{Step 6:} Finally, we rank the $m$ routes in the descending order according to their scores $C_{i}$.

\section{A Demonstrative Case Study}
 
\begin{figure}[h]
    \centering
    \includegraphics[scale=0.5]{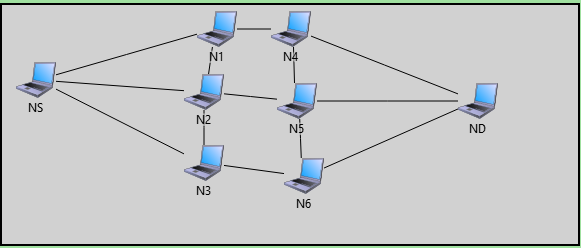}
    \caption{Network topology considered to generate the dataset.}
    \label{fig:topology}
\end{figure}

In this section, we illustrate the proposed approach on a case study.  To this aim, we considered the OMNET simulator to design the network topology described in Fig \ref{fig:topology} that connects a source node  $NS$ to a destination node $ND$. Thereof,  27 possible acyclic routes are offered. For each link, we fixed its data rate, then for each route we simulated the transmission of 1000 packets of 1 kB to compute different metrics. The dataset shown in Tab \ref{Tab-1} is thus obtained representing the 27 routes, performances of which are evaluated according to the considered metrics.

\begin{table}[h!]%
\begin{center}
{\tiny
\caption{ Data set of 27 routes with their QoS metrics.}
\label{Tab-1}
\centering

\begin{tabular}{|c|c|c|c|c|c|c|}
\hline
Route &
  \begin{tabular}[c]{@{}c@{}}Hop \\ Count\end{tabular} &
  \begin{tabular}[c]{@{}c@{}}Data rate\\ KB/s\end{tabular} &
  \begin{tabular}[c]{@{}c@{}}Packet \\ loss\\ \%\end{tabular} &
  \begin{tabular}[c]{@{}c@{}}Throughput \\ KB/s\end{tabular} &
  \begin{tabular}[c]{@{}c@{}}E2ED \\ sec\end{tabular} &
  \begin{tabular}[c]{@{}c@{}}Jitter\\ sec\end{tabular} \\ \hline
$a{1 }$ & 7 & 1571.428571 & 25.3 & 748.496994  & 0.004643 & 0.001338 \\ \hline
$a{2 }$ & 6 & 1500        & 21.7 & 785.356068  & 0.00431  & 0.001275 \\ \hline
$a{3 }$ & 5 & 1570        & 26.7 & 734.468938  & 0.00335  & 0.001363 \\ \hline
$a{4 }$ & 5 & 1600        & 14.2 & 858.858859  & 0.003349 & 0.001166 \\ \hline
$a{5 }$ & 5 & 1770        & 20.9 & 791.791792  & 0.003056 & 0.001265 \\ \hline
$a{6 }$ & 4 & 1500        & 12.5 & 877.632899  & 0.003016 & 0.001141 \\ \hline
$a{7 }$ & 5 & 1770        & 22   & 780.780781  & 0.003107 & 0.001282 \\ \hline
$a{8 }$ & 4 & 1500        & 14   & 861.723447  & 0.003067 & 0.001162 \\ \hline
$a{9 }$ & 3 & 1666.666667 & 8.1  & 919.91992   & 0.002067 & 0.001088 \\ \hline
$a{10}$ & 5 & 1750        & 17.2 & 828.828829  & 0.003138 & 0.001208 \\ \hline
$a{11}$ & 6 & 1933.333333 & 25.2 & 1123.123123 & 0.003179 & 0.000892 \\ \hline
$a{12}$ & 4 & 1937.5      & 11.5 & 1328.828828 & 0.002138 & 0.000753 \\ \hline
$a{13}$ & 4 & 1775        & 21   & 1186.186186 & 0.002279 & 0.000844 \\ \hline
$a{14}$ & 5 & 1650        & 19.1 & 810.621242  & 0.003238 & 0.001235 \\ \hline
$a{15}$ & 6 & 1708.333333 & 21.6 & 1177.177177 & 0.003571 & 0.000851 \\ \hline
$a{16}$ & 4 & 1812.5      & 10.4 & 1345.345345 & 0.002278 & 0.000744 \\ \hline
$a{17}$ & 3 & 1750        & 7.4  & 926.926927  & 0.001944 & 0.00108  \\ \hline
$a{18}$ & 4 & 2025        & 15.6 & 1562.962962 & 0.001985 & 0.000641 \\ \hline
$a{19}$ & 6 & 1666.666667 & 18.3 & 817.817818  & 0.003876 & 0.001224 \\ \hline
$a{20}$ & 7 & 1835.714286 & 24.3 & 1137.775551 & 0.003917 & 0.00088  \\ \hline
$a{21}$ & 5 & 1800        & 12.3 & 1318.136272 & 0.002876 & 0.00076  \\ \hline
$a{22}$ & 5 & 1700        & 10.9 & 1337.837837 & 0.003016 & 0.000748 \\ \hline
$a{23}$ & 5 & 1870        & 19   & 1217.434869 & 0.002723 & 0.000822 \\ \hline
$a{24}$ & 4 & 1625        & 10.5 & 895.895896  & 0.002683 & 0.001117 \\ \hline
$a{25}$ & 3 & 1616.666667 & 14.6 & 1282.282282 & 0.001874 & 0.000781 \\ \hline
$a{26}$ & 4 & 1500        & 13.9 & 861.861862  & 0.002833 & 0.001162 \\ \hline
$a{27}$ & 5 & 1600        & 14.8 & 1279.279279 & 0.003167 & 0.000783 \\ \hline
\end{tabular}
}
\end{center} 
\end{table} 

Tab \ref{tab2}, shows the different routing requirements to consider. We assume that the user application gives more importance to the E2ED whereas the hop count and the data rate are the less regarded. The application specifies also  constraints to meet for each metric.

\begin{table}[h!]%
\caption{ Criteria and  values constraints specification.}
\label{tab2}
\begin{center}
{\tiny
\begin{tabular} {|c||c|c|c|c|}
\hline
Criteria   & Nature     & Weight         & Value Range   & Value Constraints   \\ \hline \hline
Hop Count    &      Cost  &  0.05  &[3,7]         & [$\circ$,5]        \\ \hline
Data rate     &     Benefit     & 0.05  &  [1500, 2025]           & [ 1600 , $\circ$]        \\  \hline
Packet loss     &  Cost           &   0.2 & [7.4, 26.7]         &   [ $\circ$, 20 ]      \\  \hline
 Throughput    &    Benefit         &   0.2 &  [734.469, 1562.963]      &    [800, 1300]     \\  \hline
 
 End to end delay     &  Cost            & 0.3  &  [0.001874, 0.004643]         & [ 0.002, 0.004 ]        \\  \hline
Jitter     &    Cost         &       0.2     &  [0.000641, 0.001363]  &   [0.0008, 0.0012]      \\  \hline
\end{tabular}
}
\end{center}
\end{table}

\subsection{ Computing the matrix of constraints satisfaction degrees }

\begin{table}[h!]%
\begin{center}
\caption{ The Matrix of constraints satisfaction degrees.}
{\tiny
\label{Tab-3}
\centering
\begin{tabular}{|c|c|c|c|c|c|c|c|}
\hline
Route & Hop Count & Data rate & Packet loss & Throughput & \begin{tabular}[c]{@{}c@{}}End to\\  end delay\end{tabular} & Jitter & V(i) \\ \hline
$a{1}$  & 1           & 0.7171 & 0.791 & 0.8049 & 1           & 0.8466 & 0 \\  \hline
$a{2}$  & 0.5         & 0.0099  & 0.2537 & 0.9445 & 0.4821 & 0.4601 & 0 \\  \hline
$a{3}$  & 0.3333 & 0.703 & 1           & 0.7517 & 0.9994 & 1          & 0 \\  \hline
$a{4}$  & 0.3333 & 1           & 0.1299  & 1           & 0.9994 & 0.9998 & 1 \\ \hline
$a{5}$  & 0.3333 & 1           & 0.1343 & 0.9689  & 0.9994 & 0.3988 & 0 \\  \hline
$a{6}$  & 0.3333 & 0.0099  & 0.1299  & 1           & 0.9994 & 0.9998 & 0 \\  \hline
$a{7}$  & 0.3333 & 1           & 0.2985 & 0.9272 & 0.9994 & 0.5031 & 0 \\ \hline
$a{8}$  & 0.3333 & 0.0099  & 0.1299  & 1           & 0.9994 & 0.9998 & 0 \\  \hline
$a{9}$  & 0.3333 & 1           & 0.1299  & 1           & 0.9994 & 0.9998 & 1 \\ \hline
$a{10}$ & 0.3333 & 1           & 0.1299  & 1           & 0.9994 & 0.0491 & 0 \\ \hline
$a{11}$ & 0.5         & 1           & 0.7761 & 1           & 0.9994 & 0.9998 & 0\\ \hline
$a{12}$ & 0.3333 & 1           & 0.1299  & 0.8908 & 0.9994 & 0.2883 & 0 \\  \hline
$a{13}$ & 0.3333 & 1           & 0.1493 & 1           & 0.9994 & 0.9998 & 0  \\ \hline
$a{14}$ & 0.3333 & 1           & 0.1299  & 1           & 0.9994 & 0.2147 & 0 \\ \hline
$a{15}$ & 0.5         & 1           & 0.2388  & 1           & 0.9994 & 0.9998 & 0 \\ \hline
$a{16}$ & 0.3333 & 1           & 0.1299  & 0.8282 & 0.9994 & 0.3436 & 0\\ \hline
$a{17}$ & 0.3333 & 1           & 0.1299  & 1           & 0.0871 & 0.9998 & 0 \\ \hline
$a{18}$ & 0.3333 & 1           & 0.1299  & 0.0038  & 0.0233 & 0.9755 & 0\\ \hline
$a{19}$ & 0.5         & 1           & 0.1299  & 1           & 0.9994 & 0.1472 & 0 \\ \hline
$a{20}$  & 1           & 1           & 0.6418 & 1           & 0.9994 & 0.9998 & 0\\ \hline
$a{21}$ & 0.3333 & 1           & 0.1299  & 0.9313 & 0.9994 & 0.2454 & 0\\ \hline
$a{22}$  & 0.3333 & 1           & 0.1299  & 0.8567 & 0.9994 & 0.319 & 0\\ \hline
$a{23}$ & 0.3333 & 1           & 0.1299  & 1           & 0.9994 & 0.9998 & 1 \\ \hline
$a{24}$ & 0.3333 & 1           & 0.1299  & 1           & 0.9994 & 0.9998 & 1 \\ \hline
$a{25}$ & 0.3333 & 1           & 0.1299  & 1           & 0.196 & 0.1166 & 0 \\ \hline
$a{26}$ & 0.3333 & 0.0099  & 0.1299  & 1           & 0.9994 & 0.9998 & 0\\ \hline
$a{27}$ & 0.3333 & 1           & 0.1299  & 1           & 0.9994 & 0.1043 &0  \\ \hline
\end{tabular}}
\end{center}
\end{table} 

The first step consists in computing the matrix $F$ shown in Tab \ref{Tab-3} by using formulae (1) and (2). $F$ encodes the constraints satisfaction degrees that evaluates the closeness of the performance ratings of each route to meet the different constraints specified in Tab 2.  According to that (see column $V(i)$), we have only 4 routes among the 27 that are meeting all the constraints:  $a_{4}$, $a_{9}$,  $a_{23}$ and $a_{24}$.


\subsection{ Data normalization and positive closeness scores computation}

\begin{table}[h!]%
\begin{center}
\caption{ Computation of the coefficients $p_{ij}$.}
{\tiny
\label{Tab-4}
\centering
\begin{tabular}{|c|c|c|c|c|c|c|}
\hline
Route & Hop Count   & Data rate   & Packet loss & Throughput  & End to end delay & Jitter      \\ \hline
$a{1 }$    & 0.0138  & 0.0063 & 0.0437 & 0.022 & 0.0874       & 0.0417 \\ \hline
$a{2 }$    & 0.0059 & 0.0001 & 0.012 & 0.027 & 0.0391      & 0.0216 \\ \hline
$a{3 }$   & 0.0033 & 0.0062 & 0.0583 & 0.0201 & 0.063      & 0.0502 \\ \hline
$a{4 }$  & 0.0033 & 0.009 & 0.004 & 0.0313  & 0.063      & 0.0429 \\ \hline
$a{5 }$    & 0.0033 & 0.01 & 0.0061 & 0.028 & 0.0575      & 0.0186 \\ \hline
$a{6 }$    & 0.0026 & 0.0001 & 0.0035 & 0.032 & 0.0567      & 0.042 \\ \hline
$a{7 }$    & 0.0033 & 0.01 & 0.0143 & 0.0264 & 0.0584      & 0.0237 \\ \hline
$a{8 }$    & 0.0026 & 0.0001 & 0.004 & 0.0314 & 0.0577      & 0.0428 \\ \hline
$a{9 }$    & 0.002 & 0.0094 & 0.0023 & 0.0335 & 0.0389      & 0.04 \\ \hline
$a{10}$    & 0.0033 & 0.0098 & 0.0049 & 0.0302 & 0.059      & 0.0022 \\ \hline
$a{11}$   & 0.0059 & 0.0109 & 0.0427 & 0.0409 & 0.0598      & 0.0328 \\ \hline
$a{12}$    & 0.0026 & 0.0109 & 0.0033 & 0.0431 & 0.0402      & 0.008 \\ \hline
$a{13}$    & 0.0026 & 0.01 & 0.0068 & 0.0432 & 0.0429      & 0.0311 \\ \hline
$a{14}$    & 0.0033 & 0.0093 & 0.0054  & 0.0295 & 0.0609      & 0.0098   \\ \hline
$a{15}$    & 0.0059 & 0.0096 & 0.0113 & 0.0429 & 0.0672      & 0.0313 \\ \hline
$a{16}$    & 0.0026 & 0.0102 & 0.0029 & 0.0406 & 0.0428      & 0.0094 \\ \hline
$a{17}$    & 0.002 & 0.0098 & 0.0021 & 0.0338   & 0.0032       & 0.0397 \\ \hline
$a{18}$    & 0.0026 & 0.0114 & 0.0044 & 0.0002 & 0.0009      & 0.023 \\ \hline
$a{19}$    & 0.0059 & 0.0094 & 0.0052 & 0.0298  & 0.0729      & 0.0066 \\ \hline
$a{20}$    & 0.0138  & 0.0103 & 0.034 & 0.0415 & 0.0737       & 0.0324 \\ \hline
$a{21}$    & 0.0033 & 0.0101 & 0.0035 & 0.0447 & 0.0541      & 0.0069 \\ \hline
$a{22}$    & 0.0033 & 0.0096 & 0.0031 & 0.0418 & 0.0567      & 0.0088 \\ \hline
$a{23}$    & 0.0033 & 0.0105 & 0.0054 & 0.0444 & 0.0512       & 0.0302 \\ \hline
$a{24}$    & 0.0026 & 0.0091 & 0.003 & 0.0326 & 0.0505      & 0.0411 \\ \hline
$a{25}$    & 0.002 & 0.0091 & 0.0041 & 0.0467 & 0.0069      & 0.0034 \\ \hline
$a{26}$    & 0.0026 & 0.0001 & 0.0039 & 0.0314 & 0.0533      & 0.0428 \\ \hline
$a{27}$    & 0.0033 & 0.009 & 0.0042 & 0.0466 & 0.0596      & 0.003 \\ \hline
\end{tabular}
}
\end{center}
\end{table} 

The matrix $P$ shown in Tab \ref{Tab-4} reports the normalized weighted performances of each route according to the importance of each metric and the closeness to meet the constraints. Hence we determine the PIS and the NIS.
\newline
$R^+$=(0.002,	0.0114,	0.0021,	0.0467,	0.0009,	0.0022)
 and \newline
$R^-$=(0.0138,	0.00008,	0.0583,	0.0002,	0.0874,	0.0502)

\begin{table}[h!]%
\begin{center}
\caption{ Computation of the Closeness coefficients and final ranking.}
{\tiny
\label{Tab-5}
\centering
\begin{tabular}{|c||c|c||c|c||c|c|}
\hline 
       &  ISOCOV	& Hard 	& ISOCOV & Soft  & Topsis &    \\     
 Route  &  SCORE	& Ranking 	& Score & Ranking  & Score & Ranking    \\     
\hline \hline 
$a{1}$  & -0.792 & 27 & 0.208 & 27 & 0.0408 & 27 \\ \hline
$a{2}$  & -0.3894 & 12 & 0.6106 & 8  & 0.1657 & 26 \\ \hline
$a{3}$  & -0.7483 & 26 & 0.2517 & 26 & 0.2759  & 25 \\ \hline
$a{4}$  & 0.4757  & 4  & 0.4757 & 23 & 0.4552 & 16 \\ \hline
$a{5}$  & -0.4534 & 17 & 0.5466 & 15 & 0.3883 & 20 \\ \hline
$a{6}$  & -0.498 & 20 & 0.502 & 19 & 0.5326  & 14 \\ \hline
$a{7}$  & -0.4996 & 21 & 0.5004 & 20 & 0.3649 & 22 \\ \hline
$a{8}$  & -0.5063 & 22 & 0.4937 & 21 & 0.501 & 15 \\ \hline
$a{9}$  & 0.6008   & 1  & 0.6008  & 11 & 0.6913 & 8  \\ \hline
$a{10}$ & -0.4188 & 15 & 0.5812 & 13 & 0.4345 & 17 \\ \hline
$a{11}$ & -0.5824 & 24 & 0.4176 & 24 & 0.4177 & 18 \\ \hline
$a{12}$ & -0.295 & 7  & 0.705 & 3  & 0.8237 & 2  \\ \hline
$a{13}$ & -0.3791 & 11 & 0.6209 & 7  & 0.6057 & 9  \\ \hline
$a{14}$ & -0.445  & 16 & 0.555  & 14 & 0.387 & 21 \\ \hline
$a{15}$ & -0.51 & 23 & 0.49 & 22 & 0.4157 & 19 \\ \hline
$a{16}$ & -0.3176 & 8  & 0.6824 & 4  & 0.8289  & 1  \\ \hline
$a{17}$ & -0.269 & 6  & 0.731 & 2  & 0.7048 & 5  \\ \hline
$a{18}$ & -0.3236 & 9  & 0.6764 & 5  & 0.7918 & 3  \\ \hline
$a{19}$ & -0.4908 & 19 & 0.5092 & 18 & 0.2987 & 24 \\ \hline
$a{20}$ & -0.6151 & 25 & 0.3849 & 25 & 0.3255 & 23 \\ \hline
$a{21}$ & -0.3718 & 10 & 0.6282 & 6  & 0.6955 & 6  \\ \hline
$a{22}$ & -0.3929 & 14 & 0.6071 & 10 & 0.692 & 7  \\ \hline
$a{23}$ & 0.5856  & 2  & 0.5856 & 12 & 0.5902 & 12 \\ \hline
$a{24}$ & 0.5401  & 3  & 0.5401 & 16 & 0.6043 & 10 \\ \hline
$a{25}$ & -0.055  & 5  & 0.945  & 1  & 0.7707 & 4  \\ \hline
$a{26}$ & -0.4874 & 18 & 0.5126 & 17 & 0.5342 & 13 \\ \hline
$a{27}$ & -0.3928 & 13 & 0.6072 & 9  & 0.6022 & 11 \\ \hline
\end{tabular}
}
\end{center}
\end{table} 

Then, we compute the final aggregation scores  of the different routes when assuming either the constraints as mandatory or as soft.
To compare our approach, we confront our scores with those computed with TOPSIS. The latter provides the ranking when assuming no value constraints defined on the metrics. All the results are reported in Tab \ref{Tab-5}.
In terms of pure performances and according to TOPSIS, the best route is $a_{16}$ and the worst is $a_{1}$. However, when considering the value constraints on metrics, we observe according to ISOCOV that:\\
- If the constraints are mandatory: $a_{9}$ is the best route among the four routes meeting all the constraints.  \\
- If the constraints are specified as soft: $a_{25}$ is promoted as the best route and $a_{9}$ is only ranked 11th. This is because although $a_{9}$ is satisfying all the constraints, it is far to offer the best performances among the field. Besides, we can notice that $a_{1}$ is ranked as the worst route by all the methods.
 
 Consequently, ISOCOV makes it possible to provide a compromise decision between route performances and the user application requirements to meet.

\section{Conclusion}
We have proposed in this paper an approach based on using MCDM method to decide routing in presence of complex requirements that include the specification of thresholds on weighted metrics with the possibility to process them either as hard or soft constraints.
We detailed all the mathematical framework needed to model and decide the routing problem, and illustrate the approach through a case study. Future works will lead us to apply our approach on different protocols while considering additional features in the MCDM method.



\begin{thebibliography}{}

\bibitem{Bou2020}	M.R. Bouakouk, A.Abdelli, L.Mokdad:
Survey on the Cloud-IoT paradigms: Taxonomy and architectures. ISCC 2020: 1-6

\bibitem{p18} Tilwari, et al  EMBLR: A High-Performance Optimal Routing Approach for D2D Communications in Large-scale IoT 5G Network. Symmetry 2020, 12, 438. 

\bibitem{Bou2021}	M.R. Bouakouk, A. Abdelli, L. Mokdad: ODM-RPL: Optimized Dual MOP RPL. ISCC 2021: 1-6

\bibitem{p14} B. Farzaneh, A. K. Ahmed and E. Alizadeh, "MC-RPL: A New Routing Approach based on Multi-Criteria RPL for the Internet of Things," 9th IEEE ICCKE, 2019, pp. 420-425.

\bibitem{Abd2017} A.Abdelli  N.Badache,
Synchronized Transitions Preemptive Time Petri Nets: A new model towards specifying multimedia requirements. AICCSA 2006: 17-24


\bibitem{p10} S.B. Kim, et al. An Enhanced Mobility and Temperature Aware Routing Protocol through Multi-Criteria Decision Making Method in Wireless Body Area Networks. Appl. Sci. 2018, 8, 2245. 


\bibitem{Isocov1} A.Abdelli, Y. Hammal, L. Mokdad. ISOCOV: a New MCDM Method to Handle Value Constraints in Web Service Selection, IEEE ISCC, 2019.

\bibitem{Isocov2} A Abdelli, L. Mokdad, Y. Hammal
Dealing with value constraints in decision making using MCDM methods. JOCS. 44: 101154 (2020).

\bibitem{AHP} T. L. Saaty, “How to make a decision: the analytic hierarchy process,” European journal of operational research, vol. 48, no. 1, pp. 9–26, 1990.

\bibitem{TOPSIS}  C.L. Hwang, Y.J. Lai, T.Y. Liu, A new approach for multiple objective decision making, Comput. Oper. Res. 20 (9) (1993).

\bibitem{VIKOR} S. Opricovic and G. Tzeng, “Extended VIKOR method in comparison with outranking methods,” European Journal of Operational Research, vol. 178, no. 2, pp. 514–529, 2007.

\bibitem{SAW} L. Zadeh, Optimality and non-scalar-valued performance criteria, IEEE Transactions on Automatic Control 8 (1) (1963) 5960.

\bibitem{RIM} E. Cables, M. T. Lamata, and J. L. Verdegay, “RIM-reference ideal method in multicriteria decision making,” Information Sciences, vol. 337, pp. 1–10, 2016.

\bibitem{Serai2018} L. W.Serrai, et al “How to deal with QoS value constraints in MCDM based web service selection,” In Concurrency and Computation: Practice and Experience, Wiley Dec. 2019.


\bibitem{p5} A.	Al-Rubaye, Y. Yeryomin and J. Seitz, "Evaluation of MCDM-based handover decision algorithms in OMNeT++," 3rd  (IDAACS-SWS), 2016, pp. 11-14.


\bibitem{p2} A. Bose Paul, S. Biswas, S. Nandi, S. Chakraborty, MATEM: A unified framework based on trust and MCDM for assuring security, reliability and QoS in DTN routing, JNCA, Vol 104, P 1-20, 2018. 


\bibitem{p17} Bhandari, K.S.; Hosen, A.S.M.S.; Cho, G.H. CoAR: Congestion-Aware Routing Protocol for Low Power and Lossy Networks for IoT Applications. Sensors 2018, 18. 

\bibitem{p6} S. S. Bhunia, S. Roy and N. Mukherjee, "Adaptive learning assisted routing in Wireless Sensor Network using Multi Criteria Decision model," IEEE (ICACCI), 2014, pp. 2149-2154.

\bibitem{p16} V. T. Lokare, P. M. Jadhav and B. K. Ugale, "QoS based Routing using the Fuzzy TOPSIS MCDM method to enhance the performance of the MANET," IEEE 5th I2CT, 2019, pp. 1-5,

\bibitem{p9} Kim, BS., Roh, B., Ham, JH. et al. Extended OLSR and AODV based on multi-criteria decision making method. Telecommun Syst 73, 241–257 (2020). 


\bibitem{p3} B-S. Kim, et al, An enhanced geographical routing protocol based on multi-criteria decision making method in mobile ad-hoc networks, Ad Hoc Networks, Vol 103, 2020.


\bibitem{p13} Tilwari, V., Maheswar, R., Jayarajan, P. et al. MCLMR: A Multicriteria Based Multipath Routing in the Mobile Ad Hoc Networks. Wireless Pers Commun 112, 2461–2483 (2020).

\bibitem{p1} D.G Akestoridis, E. Papapetrou, A framework for the evaluation of routing protocols in opportunistic networks, Computer Communications, Vol 145, 2019, P 14-28,




\bibitem{Serai2017} W. Serrai, et al Dealing with user constraints in MCDM based web service selection. ISCC 2017: 158-163














\end{thebibliography}
\end{document}